\documentclass{article}
\usepackage[numbers,sort&compress]{natbib}
\usepackage[utf8]{inputenc}
\usepackage{authblk}
\usepackage{setspace}
\usepackage[margin=1.25in]{geometry}
\usepackage{graphicx}
\usepackage{subcaption}
\usepackage{amsmath}
\usepackage{lineno}
\usepackage{rotating}
\usepackage{tabularx} % For dynamic column adjustment
\usepackage{booktabs} % For better-looking table rules
\usepackage{algorithm}
\usepackage{algpseudocode} % For pseudocode
\usepackage{threeparttable} % For table footnotes
\usepackage{url}
\usepackage{doi}
\usepackage{array}
\newcolumntype{L}[1]{>{\raggedright\arraybackslash}p{#1}}
\newcolumntype{Y}{>{\raggedright\arraybackslash}X}

\title{Predictive Modeling, Pattern Recognition, and Spatiotemporal Representations of Plant Growth in Simulated and Controlled Environments: A Comprehensive Review}

\author[1]{Mohamed Debbagh}
\author[1]{Shangpeng Sun}
\author[1]{Mark Lefsrud}

\affil[1]{Department of Bioresource Engineering, McGill University, Montr{\'e}al, Canada.}
\date{}

\begin{document}

\maketitle

%%%%%% Abstract %%%%%%
\begin{abstract}
Accurate predictions and representations of plant growth patterns in simulated and controlled environments are important for addressing various challenges in plant phenomics research. This review explores various works on state-of-the-art predictive pattern recognition techniques, focusing on the spatiotemporal modeling of plant traits and the integration of dynamic environmental interactions. We provide a comprehensive examination of deterministic, probabilistic, and generative modeling approaches, emphasizing their applications in high-throughput phenotyping and simulation-based plant growth forecasting. Key topics include regressions and neural network-based representation models for the task of forecasting, limitations of existing experiment-based deterministic approaches, and the need for dynamic frameworks that incorporate uncertainty and evolving environmental feedback. This review surveys advances in 2D and 3D structured data representations through functional-structural plant models and conditional generative models. We offer a perspective on opportunities for future works, emphasizing the integration of domain-specific knowledge to data-driven methods, improvements to available datasets, and the implementation of these techniques toward real-world applications.
\end{abstract}

%%%%%% Main Text %%%%%%

%%%%%% Intoduction %%%%%%

\section{Introduction}\label{introduction}
With the growing prevalence of challenges in global food systems, including restricted access to fresh produce in remote areas and issues faced within traditional cultivation practices, outlooks on controlled environment agriculture (CEA) offer prospective solutions through precision methods of localized food production \cite{Mirzabaev2023,OSULLIVAN2019133,KurtBenke10.1080/15487733.2017.1394054}. Controlled environments within crop production systems provide opportunities to observe and cultivate plants with diverse characteristics, or phenotypes, making it suitable for specialized applications from trait selection for cultivar development and biopharming \cite{Halperin/10.1111/tpj.13425,HENRY2020168,langstroff2022opportunities,agronomy13030858,YANG2023102267}, to food production in extreme and resource scarce environments \cite{NGUYEN2023708,10.3389/fpls.2020.00801,10.3389/fpls.2020.00656}. Subsequently, the ability to accurately predict and represent the phenotypic expression of a plant and its trajectory at various points in its growth cycle demonstrates a fundamental step towards advancing simulation and modeling research within CEA. 

Recent developments in computational and algorithmic theories, coupled with innovations in data collection and fusion techniques, have opened new avenues for exploring plant phenotypes and growth patterns within simulated and controlled environments \cite{10.1111/tpj.14722}. These environments allow for the precise manipulation of variables such as light, temperature, humidity, and nutrient availability, enabling researchers to study the dynamic interactions that influence plant development. By leveraging techniques in predictive pattern recognition, models can be developed to accurately capture and represent plant growth behaviors.

%%%%%% Background %%%%%%
% \subsection{Background}\label{Background}
\subsection{Comparing Controlled and Open Field Environments}
To contextualize the focus of this review, it is necessary to compare the differences in observational parameters between controlled environment and open field systems. 
Open field environments, as with traditional agricultural cropping systems, are characterized by the exposure to highly variable and often uncontrollable factors including precipitation, solar conditions, climate fluctuations, and biotic stressors such as pest, pathogens, and soil microbe root interactions. 
These fluctuations introduce noise and confounding variables into experimental data, making it difficult to isolate causal relationships in plant development. 
Moreover, instrumentation methods for capturing these factors of variation (FOV) are often developed for constrained experiments, where samples are collected manually, and remain difficult to scale for application in traditional agricultural systems. Thus, outlooks within open-field research often highlight the bottlenecks in sensing and data handling which in turn limit the assessment of physiological development and crop forecasting to sparse spatiotemporal samples of FOV \cite{10.1093/jxb/erv345,WHITE2012101}.    
However, due to the desire for open environment resilient plant varieties for outdoor cropping systems, trends in field-based phenomics research have primarily focused on phenotype-genotype mapping for applications in breeding programs within open-field environments \cite{DEERY20219871989,nph.15817,ARAUS201452}. 

Controlled environments are typically found within CEA systems such as greenhouses, vertical farms, and growth chambers and are characterized by a confined growing space in which the environmental conditions are known, regulated, or fully controlled. 
These controlled environment conditions give rise to regularized observations made on plant behavior by constraining some FOVs within known parameterized boundaries.  
While not all environmental conditions can be fully isolated, key factors such as light (intensity, duration, spectrum), air and root-zone temperature, humidity, CO\textsubscript{2} levels, and nutrient delivery are routinely monitored and controlled \cite{srivani2019controlled}.
Controlled environments lend themselves well to modeling systems as instrumentation and control systems are easily integrated and necessary to maintain optimal conditions for plant growth. 
These measurements can then be parameterized to develop advanced representation models for predictive forecasting and simulation.

%\pagebreak

\subsection{Defining Simulated Environments and Plant Growth}
The phenotypic outcomes of a plant are influenced by high-dimensional interactions between genotype, environment, and management (G$\times$E$\times$M), often resulting in complex non-linear responses \cite{cooper2022predicting,MESSINA2018151,mahmood2022genotype}. 
Subsequently, the objective of controlled environment systems is to constrain and reduce the influence of the environmental dimension by maintaining known optimal conditions of various crop varieties. 
However, a second useful objective of these systems becomes apparent as advances in automated sensing and controls are implemented within CEA; The development of robust simulations and plant growth prediction models.

The term simulation can take on two meanings within this context. 
The first, is an empirical simulation where CEA is applied to mimic open-field conditions such that, the plant's behavior can be explored under controlled conditions, and the insights to be employed under real world field conditions.  
The second, is a computational simulation, or \textit{in silico} representation and quantification of plant behaviors given simulated conditions through models. 
In this review, the use of the term simulation borrows from both meanings and describes \textit{in silico} models that predict plant growth under real world conditions, which are developed by observing plant behaviors under controlled environments.

With this definition, it is also important to define plant growth within these models. 
Plant growth refers to the manifestation of plant traits, or phenotypes, over time. More specifically, the size, mass, and complexity of a plant over time, resulting from cell division, cell expansion, and differentiation processes that are regulated by internal genetic programs and modulated by external environmental factors \cite{Brukhin_Morozova_2011,alberts2002plant}. 
Quantitatively, plant growth can be represented by morphological traits such as  leaf area, height, and biomass, physiological measurements such as photosynthetic rate and transpiration, or biochemical markers, all of which vary across developmental stages and are sensitive to both spatial and temporal heterogeneity \cite{annurev-arplant-050312-120137}.
Moreover, the outputs of these simulations can have various representations from low dimensional quantitative traits to high dimensional 2D and 3D renderings and are explored further in this review.

\subsection{Background: Predictive Plant Growth Modeling}
Predictive models within plant research are often dominated by a frequentist approach to statistical inference, where empirical observations and static conditions are tested against hypotheses to determine the probable outcome through confidence intervals or point estimations. 
This statistical approach is favored within the field for its simplicity in analysis and interpretation. 
Frequentist methods, including various forms of regression modeling, provide robust tools for analyzing experimental data and have become the standard in plant research due to their well-established procedures and ease of use \cite{hoshmand2018design,bender2020statistical}. 
However, this experiment-centric approach is often limited when applied toward plant observation tasks that involve complex data structures, as with analyzing sequential and spatiotemporal images collected within the domain of high-throughput plant phenotyping (HTPP). 
Moreover, plant growth modeling and forecasting is an ill-posed problem where the trajectory of a plant's development does not have a unique solution. 
Dynamic growing conditions and stochastic elements within plant development are prevalent due to the inherent variability and uncertainty that exists in biological systems. 
For example, plant growth is influenced by numerous interacting factors, including G$\times$E$\times$M variables. Thus, methods for regularization of the problem should be explored.

\begin{figure}[t]
\centering
\includegraphics[width=\textwidth]{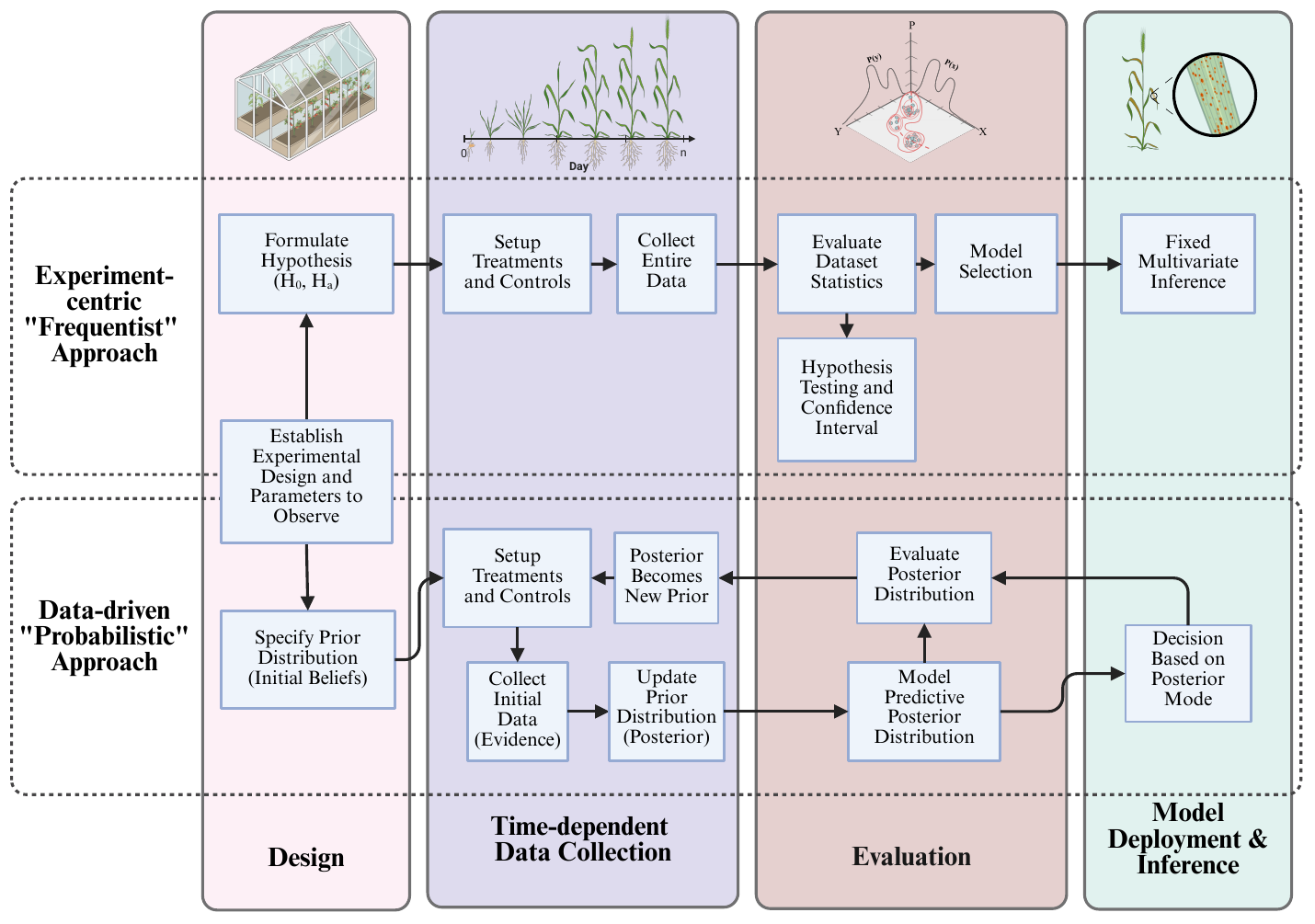}
\caption{Overview of two modeling paradigms in plant growth research. The Frequentist approach emphasizes fixed data collection, hypothesis testing, and multivariate inference within an experimental-centric framework. In contrast, the probabilistic approach employs dynamic data integration, iterative updates to prior distributions, and posterior evaluation for decision-making, enabling data-driven adaptability over time.\label{modeling paradigm}}
\end{figure}

% \pagebreak

Within the context of simulations, computational models aim to predict plant traits, such as biomass accumulation, crop yield, presence of diseases, and overall growth dynamics, during various development stages under varying environmental conditions \cite{bdcc5010002}. Some classical methods have approached this task through a combination of process models that simulate physiological principles and data-driven empirical models, typically obtained through a frequentist framework, to form various subsystems of an overarching hybrid model, offering a detailed understanding of how plants respond to different environmental stimuli at a field scale. Models such as the Crop Environment Resource Synthesis (CERES) \cite{BASSO201627} and the Decision Support System for Agrotechnology Transfer (DSSAT) have been extensively used to forecast plant growth by integrating weather data, soil conditions, and crop management practices \cite{JONES2003235, NGUESSAN2023276}. These models have proven instrumental in evaluating the potential impacts of climate change on crop productivity, guiding decision-making in traditional agriculture. However, these methods often demonstrate the various limitations of a experiment-centric approach.

The first limitation of this classical framework is that the underlying inference models are typically discriminative. The parameters of these models are treated as fixed variables that characterize boundaries used to determine the most probable outcome based on the frequencies of occurrence in sample data. This lack of flexibility in describing data distributions makes these models less effective in representing systems that involve stochastic processes, as with the various biological processes that constitute plant growth. As a result, the deterministic process models derived from these empirical outcomes are often sensitive to initial conditions and are constrained to rigid scopes of inference, making it unsuitable for distant future predictions and the synthesis of realistic data. The second limitation is that the outputs of such models are typically expressed through low-dimensional or scalar quantitative traits, such as yield, biomass, and vegetative index. While this allows for simple comparative analysis and assessments of seasonality of data points, insights toward spatial information and links between input conditions and plant structural developments are lost. The third limitation is that frequentist-based inference models do not intrinsically incorporate mechanisms to update parameters and integrate new information dynamically. This paradigm makes these models less effective for sequence analysis, as they provide no insights on prior information and treat parameters and datasets as static. In contrast, probabilistic approaches, such as those based on Bayesian inference (BI), explicitly quantify uncertainties and dynamically update with new data \cite{mackay2003information}. The static nature of the classical approach to modeling plant growth can lead to the underestimation of uncertainty, as the underlying assumptions about the "true" distribution of data may change, compounded by the cumulative effect of decisions made during earlier stages of inference in sequence analysis \cite{mcelreath2018statistical}. An overview of these dynamics between modeling paradigms are illustrated in Figure \ref{modeling paradigm}.

\subsection{Scope of this Review}
We explore these paradigms and assess the body of works available for various modeling tasks that require spatiotemporal analysis and pattern recognition of plant traits during its growth and development. 
We further propose insights to advance future works within this field and provide a perspective towards a model-centric probabilistic approach to plant growth pattern recognition and modeling through structured 2D and 3D data representations. 
This review serves as a comprehensive examination of the field of pattern recognition in plant growth.
The review is organized as follows:

\begin{itemize}
    \item We present our rational and approach for presenting the works and state our limitations. (Section \ref{Rationale})
    \item We define plant growth modeling through various temporal paradigms and examine the prominent methods explored for scalar quantitative traits. (Section \ref{sec:modeling_paradigm})
    \item We examine the application of these modeling paradigms towards 2D and 3D representations. (Section \ref{sec:structured})
    \item We explore progress made within the developing field of probabilistic and generative computer-vision that utilize image-based forecasting. (Section \ref{sec:CV})
    \item We provide a prospective towards the current state of research and propose a path for future works within the field. (Section \ref{sec:perspectives})
\end{itemize}

%%%%%% Rationale and Approach %%%%%%

\section{Rationale and Approach}\label{Rationale}
Recent state-of-the-art (SOTA) developments made within the field of HTPP have led to the establishment of data collection platforms and protocols for multimodal datasets, which include image sensor and 3D point cloud data \cite{walter2015plant,10.3389/fbioe.2020.623705,saric2022applications,plants11172199,WEN2023165626}. With the introduction of these complex datasets and multi-modal data fusion techniques, plant phenomic research has shifted towards computer vision (CV) and sensor-based methods for efficient assessment of plant traits when compared to traditional, labor-intensive approaches \cite{10.3389/fpls.2019.00508,10.1093/jxb/erz406}. Advances in CV, particularly through the use of deep learning (DL) models such as convolutional neural networks (CNN), have enabled the extraction of phenotypic traits such as leaf area, plant height, and disease symptoms from high-resolution images with good accuracy \cite{singh2016machine}. 3D imaging techniques, including structure-from-motion (SfM), stereo vision, and light detection and ranging (LiDAR) scanning, provide methods for processing spatial information to produce reconstructed 3D representation of plant architectures \cite{SUN2020195}. 

While these datasets are applied towards methods for assessing crop quality and quantifying biophysical properties throughout various stages of a plant's growth cycle, the scope of phenomics research is often limited to recognition tasks such as detection and segmentation, with limited work done on CV-based predictive forecasting. In contrast, this review focuses on work that explore the spatiotemporal components of plant phenomics research and the representation of plant traits through time.

\subsection{Literature Overview}
\begin{figure}[t]
\centering
\includegraphics[width=\textwidth]{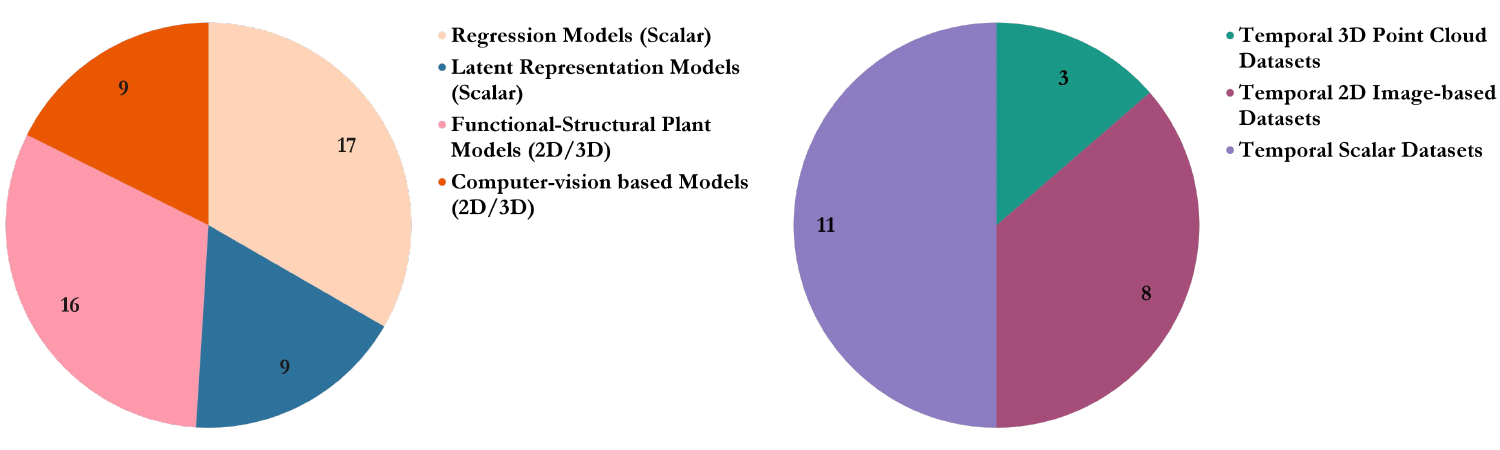}
\caption{Breakdown of publications reviewed, categorized by modeling approaches (left) and dataset types (right). The figure highlights the distribution of works across regression models, latent representation models, functional-structural plant models, and computer vision-based techniques, as well as their focus on temporal scalar, 2D image-based, and 3D point cloud datasets.\label{publication overview}}
\end{figure}

A rigorous screening process was applied to articles indexed within the following publication databases: SCOPUS, Web of Science (WOS), PubMed, and Google Scholar. 
A total of 51 method papers on plant growth modeling and 22 temporal dataset papers were selected and reviewed (Figure \ref{publication overview}). 
These papers include articles published between the years, 2015-2025, with a particular focus on assessing plant traits over time, where temporal analysis is a key component to the method or dataset.
Foundational works predating 2015, such as standard numerical crop models, are referenced for context, but are not explored in depth as they are beyond the scope of this review.
Articles in which traits are observed sparsely over time, for example traits observed once at the beginning of an experiment and once after harvest, are excluded from this review, opting for consistent, relatively high temporal resolution tracking of traits over time.

\subsection{Plant Growth Representations and Datasets}

A repository of recently published datasets on various temporal plant growth dynamics was compiled for this review and is shown in Table \ref{dataset list}. This repository establishes several representations of observable plant traits which can be categorized into broader segments and provides insights towards the current focus of plant phenomics research. These data representations include: 3D representations of plant structures, typically expressed through 3D point clouds; 2D representations of plant structures expressed through images; and finally, the most common amongst a broader spectrum of classical methods, is 1D representation of various traits expressed through quantitative scalar data. 

\subsection{Limitations}
This review highlight works that have contributed significantly towards advanced techniques in pattern recognition by providing deeper insights into the mathematical frameworks or, in the case of more traditional forecasting methods, are recent and representative of the fundamental principles with some novel insights discussed.
Due to the exploratory nature of this review of novel and SOTA approaches within this developing field, the availability of works and datasets that cover longitudinal predictive forecasting and simulated growth representations and meet this criterion is limited. 
We have omitted some works that simply use models already established by other works, and thus the number of works available for a full systemic review is not possible. 
Moreover, this review is constrained to published research and datasets, which may lead to an inherent bias towards studies and datasets that have achieved publication rather than those that may be equally relevant but remain in preprint stages, unpublished, or private.

\begin{sidewaystable}
\caption{Comprehensive list of temporal plant research datasets, published between 2015 and 2025, capturing various growth dynamics and traits. \label{dataset list}}
\scriptsize % Shrinks the font size for large tables
\begin{tabularx}{\textwidth}{@{\extracolsep{\fill}}p{0.15\textwidth}p{0.2\textwidth}X X@{\extracolsep{\fill}}}
\toprule
\textbf{Dataset} & \textbf{Crop} & \textbf{Traits} & \textbf{Data Type} \\
\midrule
Sun et al. (Soybean-MVS) \cite{Sun2023soybeanmvs} & Soybean (\textit{Glycine max var. DN251, DN252, DN253, HN48, HN51}) & Segmented plant structure (leaf shape, stem, branches) & Point cloud (3D) \\
\hline
Schunck et al. (Pheno4D) \cite{10.1371/journal.pone.0256340} & Maize (\textit{Zea mays}), Tomato (\textit{Solanum lycopersicum}) & Segmented plant structure (leaf shape, stem, branches) & Point cloud (3D) \\
\hline
Sultan et al. (Shoots and Roots) \cite{doi:10.34133/2020/2073723} & Maize (\textit{Zea mays}), Tomato (\textit{Solanum lycopersicum}), rice (Azucena, Bala), Arabidopsis (\textit{Arabidopsis thaliana}) & Root system traits and structure & Point cloud (3D) \\
\hline
Uchiyama et al. (KOMATSUNA) \cite{Uchiyama2017Komatsuna} & Komatsuna (\textit{Brassica rapa var. perviridis}) & Plant structure & Multi-view RGBD images and annotated masks (2D/3D) \\
\hline
Jayasuriya et al. \cite{JAYASURIYA2024110735} & Capsicum (\textit{Capsicum annuum L.}) & Plant structure & RGBD images (2D/3D) \\
\hline
Kierdorf et al. (GrowliFlower) \cite{Uchiyama2017Komatsuna} & Cauliflower (\textit{Brassica oleracea var. botrytis}) & Plant structure & RGB and multi-spectral images (2D) \\
\hline
Shadrin et al. \cite{Shadrin2018InstanceSF} & Lettuce (\textit{Lactuca sativa}) & Leaf growth and leaf segmentation & Top-down RGB images and annotated masks (2D), and environmental sensor readings (tabular) \\
\hline
Genze et al. \cite{genze2024manually} & Maize (\textit{Zea mays L.}), Sorghum (\textit{Sorghum bicolor}) & Plant structure & Top-down RGB images and annotated masks (2D) \\
\hline
Thoday-Kennedy et al. \cite{THODAYKENNEDY2023108787} & Safflower (\textit{Carthamus tinctorius L.}) & Plant morphology and salt response & Multi-view RGB images (2D) \\
\hline
Lärm et al. \cite{larm2023multi} & Maize (\textit{Zea mays}), Wheat (\textit{Triticum aestivum}) & Root system traits and structure & RGB images and ground-penetrating radar map (2D), soil sensor data (tabular) \\
\hline
Parmentier et al. \cite{essd-13-3593-2021} & General vegetation in Arctic ecosystems & Vegetative phenology and climate interactions & RGB images (2D) and environmental vegetative index data (tabular) \\
\hline
Craig et al. \cite{craig2022hyperspectral} & Maize (\textit{Zea mays L.}) & Various agronomic traits and vegetative index & Scalar (tabular) \\
\hline
D{\'e}pigny et al. (Plantain-Optim) \cite{DEPIGNY2018671} & Plantain (\textit{Musa acuminata × Musa balbisiana}) & Various agronomic traits and biomass & Scalar (tabular) \\
\hline
Ngaba et al. \cite{NGABA2023109244} & Sugarcane (\textit{Saccharum officinarum}) & Various agronomic traits and biomass & Scalar (tabular) \\
\hline
Newman et al. \cite{newman2021multiple} & Multiple field-crop species & Long-term agronomic traits and biomass & Scalar (tabular) \\
\hline
Campbell et al. \cite{CAMPBELL2021107600} & Maize (\textit{Zea mays}) & Photosynthetic function and vegetative index & Scalar (tabular) \\
\hline
Manjarrez-Sanchez et al. \cite{MANJARREZSANCHEZ2020106274} & Greenhouse Tomato (\textit{Lycopersicum esculentum}) & Photosynthetic function and agronomic traits & Scalar (tabular) \\
\hline
Niu et al. \cite{essd-14-2851-2022} & Maize (\textit{Zea mays}) & Long-term phenological and climate interactions & Scalar (tabular) \\
\hline
Zhu et al. \cite{essd-16-277-2024} & Multiple field-crop species & Long-term phenological and climate interactions & Scalar (tabular) \\
\hline
Mergner et al. \cite{mergner2020proteomic} & Arabidopsis (\textit{Arabidopsis thaliana}) & Gene expression & Scalar (tabular) \\
\hline
Castro et al. \cite{CASTRO2020105834} & Camu-camu (\textit{Myrciaria dubia}) & Gene expression & Scalar (tabular) \\
\hline
Peng et al. \cite{peng2024transcriptome} & Peanut (\textit{Arachis hypogaea}) & Gene expression & Scalar (tabular) \\
\bottomrule
\end{tabularx}
\end{sidewaystable}

\section{Temporal Modeling Paradigms of Quantitative Traits}
\label{sec:modeling_paradigm}
The abstraction of various plant biophysical properties into 1D quantitative traits provides a way to produce model parameters with functional meaning for a wide application of plant processes. Within plant phenomics, these applications can range from the molecular level with gene-protein expressions and phenotype mapping; the plant level in which we observe various plant-environment interactions; and the phenological level, where cyclical patterns of vegetative traits over a population within an ecosystem are mapped. In typical HTPP methods, scalar quantitative traits are inferred from high dimensional data such as RGB and multi-spectral images collected over time \cite{GE2016625,anche2023scalable,anche2020temporal,sun2018field}. These low dimensional quantitative traits can be assessed through time and represented through various temporal modeling paradigms.

\subsection{Regression Models}

A common approach to simulate and predict the trajectory of these traits through time is through some form of regression-based inference model. Much of the works covered in this review describe methods for pattern recognition of plant growth that adhere to this framework. Phenotypic traits, denoted as $y$, can be expressed by explicitly fitting one of various forms of regressions (Equations \ref{eqn:polynomial}-\ref{eqn:logit}). We can define the configuration of independent parameters that contribute to the expression of these traits, as $X = \{ x_i \mid i \in \{1,2, \ldots, N\}\}$ for $N$ parameters. For instance, $y$ can represent a phenotypic trait such as plant height or leaf area, and $x_i$ represents the expression level of gene $i$. In order to model these effects continuously through time, $t$, a generic regression model is given by Equation \ref{eqn:regress}. This regression approach provides an empirical basis for fitting discrete observation data and determining the fixed components of models such as regression and spline coefficients, $\beta$ and $\gamma$, respectively. Discrepancies between observed data and modeled outputs can be described as noise through error term, $\epsilon$.

%\pagebreak 

\begin{align}\label{eqn:regress}
y(t) = f(t) + \epsilon(t)
\end{align}

\begin{align}\label{eqn:polynomial}
    % &f(t) = \beta_0 + \beta_1t && \text{(Linear)} \\ \label{eqn:polynomial}
    &f(t) = \sum_{i=0}^{n} \beta_{i} t^{i}&& \text{(Polynomial)} \\ \label{eqn:spline}
    &f(t) = \beta_0 + \beta_1t  + \sum_{j=1}^{K} \gamma_j S_j(t) && \text{(Spline)} \\ \label{eqn:logit}
    &f(t) = \frac{1}{1 + e^{-(\beta_0 + \beta_1t)}} && \text{(Logistic)}
\end{align}

The most common of these include non-linear regressions models that capture the typical patterns seen within plant growth traits. Following the conventions of a frequentist approach, sources of variation amongst growth patterns are hypothesized and made explicit before the collection of data. Non-experimental parameters, such as environmental and spatial variations, are standardized or corrected. Two works, Perez et al. \cite{Perez-Valencia2022} and Kar et al.\cite{SpaTemHTP}, have proposed the implementation of spatial analysis of trials with splines (SPATS) for correcting spatial variations amongst experimental groups as a data pre-processing step before performing temporal modeling. The fixed but unknown statistics that describe the data are obtained empirically to fit the regression function for longitudinal trait predictions. The most prevalent implementations for modeling non-linear time-series patterns of scalar growth traits are spline-based forms, with common application of the model demonstrated in works such as Momen et al. \cite{Mehdi10.1534/g3.119.400346} and Pauli et al. \cite{Duke10.1534/g3.115.023515}. Polynomial forms, demonstrated in works such as Baba et al. \cite{Toshimi10.1371/journal.pone.0228118} and Morales et al. \cite{Morales10.1093/genetics/iyae037}, account for the majority of regressions used for longitudinal trait prediction.

\subsubsection{Separating Effects using Hierarchical Models}
To disentangle the individual and compounding effects of various factors on plant growth trajectories, a hierarchical regression structure can be employed to represent plant traits for different levels of variation. This can be particularly useful for complex systems representing plant growth that is determined by multiple cumulative factors. For instance, plant traits such as height or leaf area can be modeled as responses to a combination of fixed and random effects. Fixed effects capture the consistent factors, such as the overall population-wide trends in growth, while random effects model the variation at each hierarchical level including genotype-specific effects and individual plant-specific variations. The random effects can vary across time, capturing the longitudinal nature of plant growth, while still maintaining a structure that allows for partial pooling of information across the hierarchy.

\begin{figure}[t]
\centering
\includegraphics[width=\textwidth]{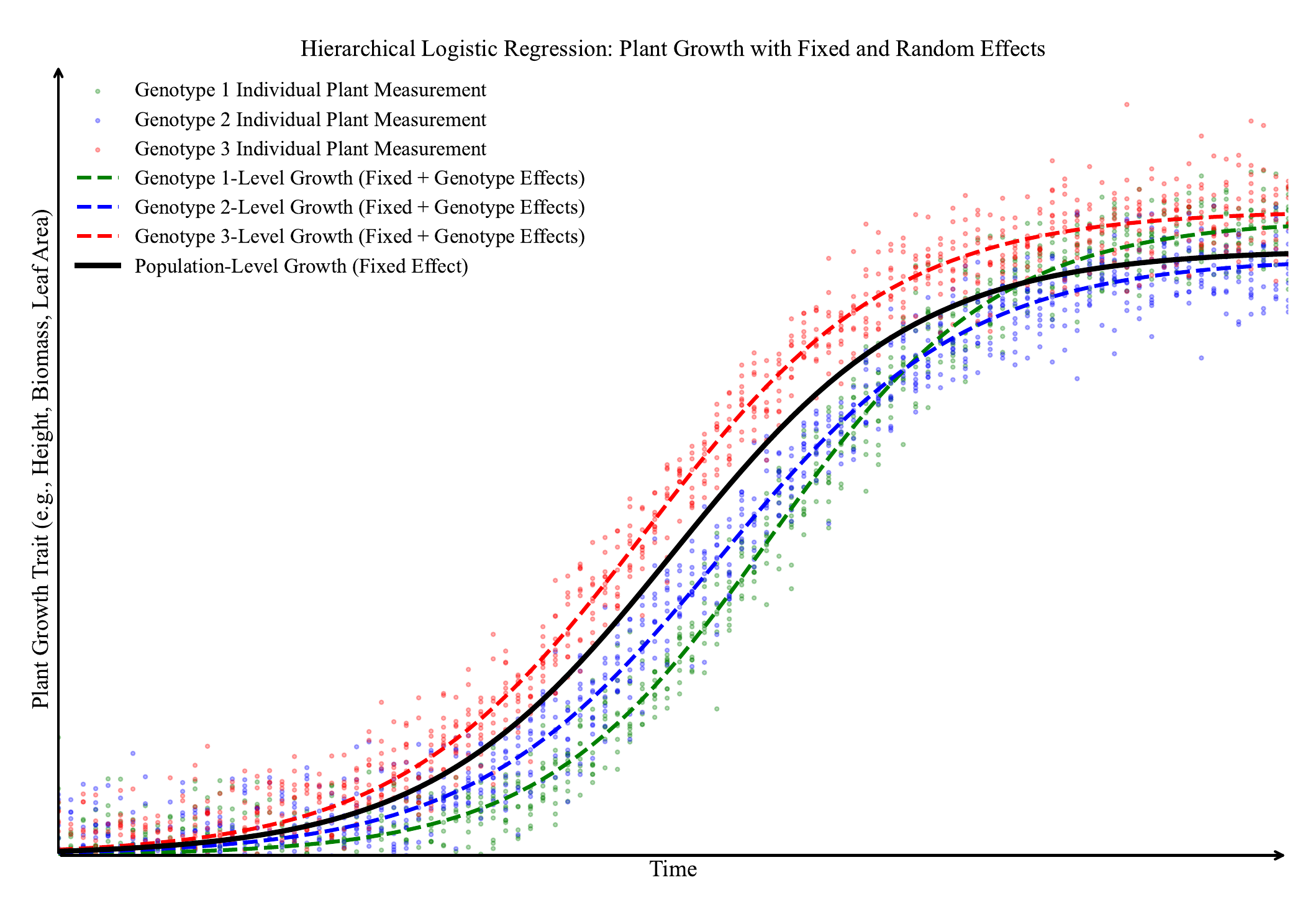}
\caption{Hierarchical logistic regression modeling of typical plant growth with fixed and random effects. This hierarchical approach highlights variability at multiple levels, with the logistic model capturing both population-wide trends and genotype-specific and plant-specific variations in plant traits such as height, biomass, or leaf area.\label{hierarchical_reg}}
\end{figure}

One such application of this is demonstrated by Equation \ref{eqn:hierarichal} and shown in Figure \ref{hierarchical_reg}. A hierarchical logistic regression is employed to represent plant growth at the population level, genotype level, and individual plant level. A generic form logistic transformation (Equation \ref{eqn:logit}) can be applied towards the compounding fixed population effects, $f_p$, the random effects at the genotype level, $f_{pg}(t)$, and the random effects of each individual plants, $f_{pgi}(t)$, to describe the development of plant traits over time, $y(t)$. In this example $\beta$, $L$, and $t_0$ in Equations \ref{eqn:fixedpop}-\ref{eqn:randindv} are model parameters to be determined in order to fit our regressions to corresponding categorized data. 

\begin{align}\label{eqn:hierarichal}
\log\left(\frac{y_{ijt}(t)}{L_{ij} - y_{ijt}(t)}\right) = f_p(t) + f_{pg}(t) + f_{pgi}(t) + \epsilon_{ijt},\:  \epsilon_{ijt} \sim \mathcal{N}(0, \sigma^2)
\end{align}

{\noindent \begin{align}\label{eqn:fixedpop}
    &f_p(t) = \beta_{p,0} + \beta_{p,1} \cdot (t - t_0) && \text{(Fixed Population Effects)}\\ \label{eqn:randgen}
    &f_{pg}(t) = \beta_{pg,0_{i}} + \beta_{pg,1_{i}} \cdot (t - t_0)&& \text{(Random Genotype Effects)} \\ \label{eqn:randindv}
    &f_{pgi}(t) = \beta_{pgi,0_{ij}} + \beta_{pgi,1_{ij}} \cdot (t - t_0) && \text{(Random Indvidual Plant Effects)}
\end{align}}

When modeling plant traits over time, fixed effects are commonly defined by a logistic regression to capture typical growth curve patterns and to distinguish temporal trends between categorical groups, such as genotypes. A typical implementation of such models is demonstrated in works such as Baker et al. \cite {Baker2018Baysian}, in which the effects of genotypes are separated using Bayesian routines to estimate logistic parameters. Certain instances may require the disentanglement of complex effects between groups. In such cases, the mapping of these relationships require that regression models capture higher degrees of non-linearity. In Perez et al. \cite{perez2024one}, a P-spline-based hierarchical structure (Equation \ref{eqn:spline}) is employed to model the effects not just at the population and genotype level but also the effects between individual plants, experimental groups and spatio-temporal variations. This approach allows for the simultaneous representation of growth curves at various levels of variability and the incorporation of shared parameters amongst each group effects on plant growth. 

\subsubsection{How Controlled Environments Affect Fixed Variables in Regression}
Temporal regression models for assessing longitudinal quantitative traits are typically developed under open-field conditions, which results in higher variability in observed outcomes and limit predictions made under dynamic conditions. 
One might want to separate these environmental effects as fixed variables to limit this variability for better performance on inference. 
While, controlled environment systems can regularize these conditions, the strict control of these factors can also introduce certain modeling constraints and affect how fixed effects are interpreted in temporal regression models.
Specifically, the fixed variables in a regression model may become conflated with the uniform environmental conditions of the controlled environment. 
For example, plant height trajectories measured under identical light and nutrient regimes may exhibit reduced variability, 
masking G$\times$E interactions that would otherwise emerge in field conditions \cite{teressa2021multi}. 
As a result, the fitted regression parameters may overrepresent the influence of genotype or treatment effects, while underestimating environmental contributions.

Moreover, fixed effect parameters often capture the average trait progression under a narrow set of environmental conditions. 
This introduces a risk of overfitting to the specific experimental regime and limits the generalizability of the model to more variable or open-field conditions \cite{hawinkel2022spatial}. 
The deterministic nature of environmental control may also induce collinearity amongst time-dependent variables, making it difficult to separate the temporal effect of a fixed factor from its environmental context \cite{dormann2013collinearity}.

\subsubsection{Pitfalls of Regression-based Temporal Inference}
\label{pitfalls of regression}
Modeling plant growth through regressions provides meaningful interpretations of trends on labeled data points. 
However, temporal inferences made from these models suffer from the pitfalls of a post-hoc analysis. 
Following the typical experiment-centric modeling pipeline (Figure \ref{modeling paradigm}), inference models are selected after data has been collected and are typically implemented to explain trends once experimental groups are found to be statistically significant. 
As such, various regression forms and their hyper-parameters are selected once this data is revealed in order to achieve best fit. 
This post-hoc selection provides an inherent risk of overfitting due to the lack of a priori model constraints, leading to interpretations that may capture transient or idiosyncratic patterns rather than robust, underlying temporal trends \cite{babyak2004you,prakash2023temporal}. 
As a result, post-hoc regression models lack the generalizability necessary to predict future data accurately, particularly in time-series with complex dependencies or irregular periodicity \cite{hong2018overfitting,fithian2014optimal}.

Another pitfall involves the interpretability of the regression parameters. Temporal inferences often rely on the assumption that parameters estimated from the data capture the true characteristics of the underlying temporal process. However, in the absence of predefined model assumptions, these parameters can be sensitive to random variations or transient anomalies in the dataset. This sensitivity can result in misleading estimates of trend strength, direction, or duration, especially in nonlinear regression models, which are more prone to parameter instability when applied post hoc \cite{gutenkunst2007universally}. Consequently, regression-based temporal models may yield unreliable estimates that lack external validity, thereby compromising the inference drawn from these analyses.

Furthermore, temporal data often contain complex dependencies due to autocorrelation, seasonal effects, or multi-modal factors. When regression models are employed in a post-hoc framework, these dependencies are frequently under accounted for, leading to temporal confounding. Such confounding poses a significant challenge as it obscures the causative factors underlying observed trends and further limits the model's applicability beyond the studied period \cite{shumway2000time}. Works such as Nesteruk et al. \cite{Sergey9152399} demonstrate the use of autoregressive model improved predictions when compared to other non-correlated regressions.

%%%%%% Latent Representation Models %%%%%%

\subsection{Latent Representation Models}
Quantitative traits can also be assessed through time by representing their complex temporal dependencies through some reduced latent representations\footnote{Latent representations refer to reduced-dimensional, abstract encodings of quantitative trait data that capture temporal dynamics (time dimension) and underlying structure.}, or hidden state. 
These representations can be applied as a transformation on sequential inputs to derive predictive outputs. In this section we explore various methods for extracting latent representation from the temporal dependencies between scalar quantitative traits and the models used to perform sequence-based inference from these hidden states.

\subsubsection{Dimensionality Reduction through Principal Component Analysis}
Methods like those employed by Adak et al. \cite{1jkae092FPCA}, aim to extract key features from time-series data that correspond to a reduced representation of growth rates using principal component analysis (PCA). In this context, PCA is a pattern recognition technique that captures the temporal components of the evolution of traits, such as key moments found within vegetative indices (VI) of plants of differing genotypes. These reduced representations effectively act as growth-rate phenotypes, embodying the entirety of a plant's growth cycle and can be incorporated as an independent variable to improve accuracy of genomic and phenomic prediction models \cite{bimj.202000315}. However, meaningful interpretations of these phenotypes, especially for simulation purposes, are lost as these representations are abstracted into a reduced latent form. Interpretations of these representations back to the time domain for inference requires a decoding process. We explore such a process for decoding a latent representation in Section \ref{rnn}.

%\pagebreak

\subsubsection{Neural Network-based Models}
More recently, with advancements in computational systems and optimization algorithms, NNs have been employed in models to perform powerful opaque mappings between inputs and outputs. As such, they have expanded the range of tools available for dimensionality reduction and feature extraction. While some early implementations of simple feed forward DL systems are employed for pattern recognition, they often perform inference on fixed time windows which result in the same pitfalls of regression systems described in Section \ref{pitfalls of regression} \cite{hewamalage2023forecast}. For instance, Matsumura et al. \cite{MATSUMURA2015} employs a multi-layer perceptron (MLP) with inputs of early season parameters in the form of a fixed window to forecast end of season outputs. While this approach can capture relationships within the window, it fails to incorporate longer-term temporal dependencies beyond the window itself. This results in models that are limited in their ability to adapt to changes outside the fixed period, potentially leading to overfitting to short-term patterns and reduced generalizability to future time steps or different seasons.

\begin{figure}[t]
\centering
\includegraphics[width=.85\textwidth]{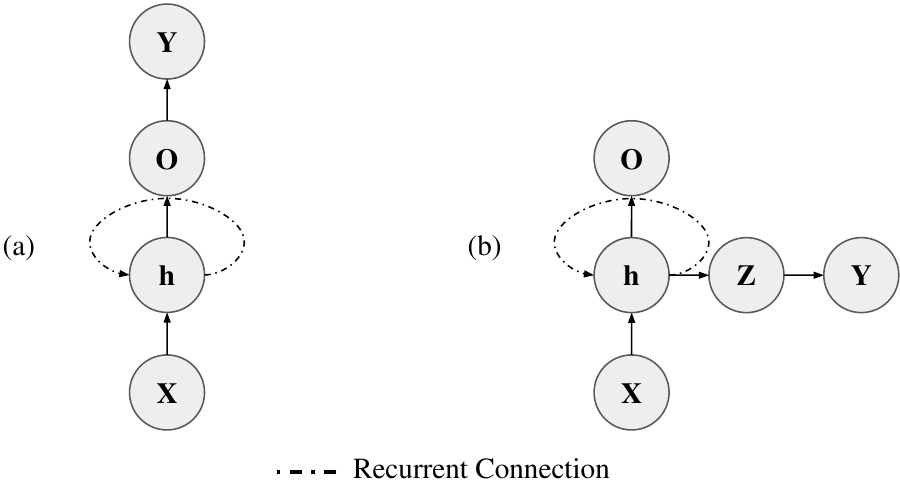}
\caption{Directed graph models (folded) of the typical recurrent neural network (RNN) pipeline for sequence prediction and encoding. 
(a) Shows the sequence prediction task, where inputs $X$ are processed through recurrent connections to generate hidden states $h$, leading to intermediate outputs $O$ and a final output $Y$. This demonstrates how RNNs can predict each subsequent element in the sequence based on prior context.(b) Shows a sequence encoding process, where $X$ is processed into a final encoded latent variable $Z$, capturing the entire sequence’s context in a single vector. $Z$ can be decoded to obtain $Y$.\label{rnn_use}}
\end{figure}

\paragraph{Recurrent Neural Networks}
\label{rnn}
To resolve this issue of fixed windows, sequence-based NNs, such as recurrent neural networks (RNN) and transformers, have been the focus of recent works to assess traits through time. Two main implementations of these networks include direct sequence predictions and latent sequence encoding (Figure \ref{rnn_use}). The former utilizes NNs for sequence generation tasks such as time-series forecasting. In the context of RNNs, prior sequential information is defined by the weights of an NN and represented as a hidden state, $h$, which embodies all cumulative prior sequential information. This $h$ is used to transform the next given input, $X_t$, to predict the next element in the sequence, $Y_{t+1}$ (Figure \ref{rnn_use}(a)). Various recent works within plant yield forecasting implement a subset of RNNs known as long short term memory (LSTM) networks to predict the overall real-time trajectory of plant growth traits \cite{lee2021development,alhnaity2019using}.

The second implantation of these NNs aims to capture and represent temporal relationships through a fixed latent variable, $Z$ (Figure \ref{rnn_use}(b)). $Z$ is obtained by processing sequential information through an RNN and is subsequently derived from $h$, also typically defined in the form of the weights of a MLP between the two states. This representation captures the entirety of the sequence as a context driven conditional and can be decoded to perform inference on various tasks. One such application is demonstrated by Gong et al. \cite{s21134537}, in which end of harvest yield is forecasted from a latent embedding of sequential environmental parameters through a temporal CNN decoder. Moreover, the decoding process can also be applied to task in which it is useful to predict future sequences of the same trait within a sequence-to-sequence approach. Works such as Bashar et al. \cite{ALHNAITY202135}, show the utility of such approach using a LSTM-based networks as both a sequence encoder to capture significant temporal relationships within plant growth development into a low dimensional embedding and as a decoder to obtain the next sequence of predicted outcomes from these embeddings. 

% \pagebreak

\paragraph{Transformers and Attention Mechanisms}
It is worth noting the applicability of transformers and attention mechanisms for analyzing sequential data similar to RNNs. However, unlike RNNs, transformers leverage a self-attention mechanism to process entire sequences in parallel, enabling them to capture long-range dependencies effectively as each element directly attends to all others \cite{transformer}. Although few studies have applied these architectures towards temporal plant trait modeling, their ability to weigh the importance of different time steps dynamically demonstrates their potential for analyzing plant growth patterns influenced by both recent and distant environmental factors. This adaptability suggests significant opportunity for advancing plant growth forecasting, particularly for complex, multi-modal time-series data.

\subsubsection{State-Space Models}
State-space models (SSM) define a broader categorization of sequence processing approaches similar to RNNs. However, rather than learning an implicit hidden state, which captures prior information as memory, SSMs are defined by an explicit transition state model and observation model. These models can be expressed flexibly between deterministic functions for dynamical systems or as probabilistic methods for describing complex stochastic processes (Table \ref{ssm_tab}). The transition state model defines how latent states, $z_{t+1}$, evolve over time, and their dependence with the prior input, $x_t$, and latent state, $z_t$. The observation model relates the latent state to observed variables, $y_t$. SSMs assume a Markov property, in which the future state of a system depends only on its current state. As such, the process has no memory beyond the current state as with RNNs. This Markov property can be relaxed with the inclusion of memory states of prior sequences in the transition state model (Section \ref{generative process}).

\begin{table}[h]
\centering
\begin{threeparttable} % Wrap the table in threeparttable for footnotes
\caption{Comparison between the components of deterministic and probabilistic state-space models.}\label{ssm_tab}
\begin{tabular*}{\textwidth}{@{\extracolsep\fill}lcc}
\toprule%
& \textbf{Deterministic SSM} & \textbf{Probabilistic SSM} \\
\midrule
State Transition Model & $\mathbf{z}_{t+1} = f_\theta(\mathbf{z}_t, \mathbf{x}_t) + \eta_t$ & $\mathbf{z}_{t+1} \sim p_\theta(\mathbf{z}_{t+1} | \mathbf{z}_t, \mathbf{x}_t)$  \\
Process Noise  & $\eta_t \approx 0$ &  $\eta_t \sim \mathcal{N}(0, \Sigma_{\eta})$   \\
\midrule
Observation Model\textsuperscript{\textdagger}  & $\mathbf{y}_t = g_\theta(\mathbf{z}_t) + \xi_t$ & $\mathbf{y}_t \sim p_\theta(\mathbf{y}_t | \mathbf{z}_t)$ \\
Observation Noise\textsuperscript{\textdagger}  & $\xi_t \approx 0$ & $\xi_t \sim \mathcal{N}(0, \Sigma_{\xi})$ \\
\bottomrule
\end{tabular*}
\begin{tablenotes}
\item[\textdagger] \small{Models may introduce stochasticity to a core deterministic functional. In such cases, process and observation noise do not equal 0.}
\end{tablenotes}
\end{threeparttable}
\end{table}

% \pagebreak

Within the context of plant growth modeling and phenological pattern recognition, SSMs present an approach that captures the gradual changes in traits by modeling transitions within latent growth states \cite{2021_AugerMethe,9129053}. This approach has the advantage of continuously updating estimates of latent variables based on context driven observations such as environmental factors. For instance, Viaud et al. \cite{viaud2022full} proposes that complex plant growth dynamics, including the variability from noisy environmental conditions can be better represented through a classical SSM framework, when compared to traditional models. More specifically, BI can be applied to estimate the parameters, $\theta$, of a hidden Markov model (HMM), through a particle MCMC method. As such the stochastic elements of plant growth can be captured through a probabilistic latent transition state. In recent works, NNs have been employed as the transition state models to capture complex non-linear relationships between environmental and plant trait development. Shibata et al. \cite{2020_4261965} introduces a deep SSM designed to predict sugar content in greenhouse tomatoes from inputs representing environmental conditions. In this work it is suggested that a semi-surpervised approach can be taken and paired with reinforcement learning for dynamic control tasks in CEA, where optimal decisions must be made in response to changing conditions.

\subsubsection{Pitfalls of Representation Models}
\label{pitfallsrepresentation}
Assessing plant growth patterns through latent representations provides a way to capture the complex long-term temporal dynamics of trait development. 
The hidden and latent states, particularly of RNNs and SSMs, determine the transformations on prior states, which can be used to interpret how the system will change through time. 
As such, these methods improve upon classic regression systems by providing a context driven inference paradigm that resolves various issues of post-hoc analysis and allows for real-time predictions. 
However, the abstraction of these temporal representations into latent forms results in challenges of interpretability \cite{lipton2018mythos,o2025interpretability}. 
Methods that inspect inner network structures reveal that even mechanistic dissection of latent variables often yields abstract or semantically opaque units \cite{rauker2023toward}.

As the prior states are transformed into latent or hidden states, the connection between the model inputs and the output space can become unclear. 
For instance, PCA reduces temporal data to principal components, but these components often lack a direct explainable meaning, making it difficult to infer how individual traits evolve over time directly and require some additional mapping to outputs \cite{jolliffe2016principal}. 
This challenge is further exacerbated if mappings are performed through NN-based models, where latent and output states are learned through complex nonlinear weight transformations, leading to opaque predictions that prevent practical interpretations. Moreover, determining the inverse mapping of the model solution is not directly possible. For example, given the outcome of a plant's trait at a particular time, we cannot determine the input conditions and prior states that gave rise to the outcome directly, especially if latent states are determined through NNs or other highly abstracted models.

\paragraph{Integrating Domain-Specific Knowledge}

Future work may focus on the integration of domain-specific knowledge within these pattern recognition paradigms, especially in cases where meaningful interpretation of processes is important. These knowledge-driven models typically define various plant processes based on biological principles such as photosynthesis, respiration, and growth dynamics through mechanistic models \cite{CASSIA}. Xing-Rong et al. \cite{FAN2015363} explores a combined data-driven and knowledge-based approach to resolve issues of strict calibration constraints, and incomplete representations of complex plant dynamics that result from purely mechanistic models. Much of these hybrid models are early in their conception, utilizing simple data-driven models, such as regressions or simple MLPs, to account for variability of outputs or as direct inputs, as an estimator, to a mechanistic model. Consequently, multiple models are often employed as separate components and consolidated through a two-step process; one step to obtain the data driven term, and the second step for the mechanistic output. This approach has the disadvantage of higher computational cost and potential loss of capturing hierarchical or interdependent effects, especially if training processes are separated. Thus, there are significant potential for further contributions of advanced data-driven techniques to improve methods within a hybrid paradigm \cite{shi2025deep}.

\subsection{Comparisons of Temporal Models}

Table~\ref{tab:summaryTemp} provides a comparative summary of the temporal modeling paradigms explored in this section for assessing longitudinal quantitative traits. Each method varies in its data requirements, ability to model uncertainty, interpretability of its parameters, and its typical use cases in plant phenomics. Regression-based models remain the most interpretable and widely applied due to their simplicity and empirical grounding, particularly in structured experimental setups. However, they often lack the flexibility to model complex temporal dependencies and are prone to post-hoc inference limitations.

Latent representation methods, including PCA and NN-based models, offer powerful tools for capturing abstract patterns and nonlinear dynamics, but at the cost of interpretability. Probabilistic models such as SSMs provide a middle ground by explicitly modeling noise and latent transitions while allowing for context-aware updates. Hybrid approaches that combine data-driven inference with mechanistic plant models may offer a promising direction for advancing interpretability and generalization across varied environmental and biological conditions.

\begin{table}[h]
    \centering
    \begin{threeparttable}
    \caption{Comparison of Temporal Models for Assessing Longitudinal Quantitative Traits}
    \label{tab:summaryTemp}
    \scriptsize
    \begin{tabularx}{\textwidth}{@{\extracolsep{\fill}}L{0.14\textwidth}L{0.18\textwidth}L{0.21\textwidth}L{0.17\textwidth}Y@{}}
    \toprule
    \textbf{Model} & \textbf{Data Requirements} & \textbf{Uncertainty Handling} & \textbf{Interpretability} & \textbf{Application} \\
    \midrule
    \textbf{Regression}\textsuperscript{\textdagger}  & Low; structured, independent data & Not modeled during inference, but handled through confidence intervals & High (coefficients are interpretable) & Growth trend analysis; treatment comparisons \\
    \addlinespace
    \textbf{Hierarchical Regression} & Moderate; nested or multi-level data & Explicit modeling via random effects & Moderate to high & Multi-environment trials; genotype $\times$ environment modeling \\
    \addlinespace
    \textbf{Feedforward NN (MLP)}\textsuperscript{\textdagger}  & High; labeled training data & Not explicitly handled & Low; black-box weights & Trait prediction; nonlinear phenotype modeling \\
    \addlinespace
    \textbf{Sequence RNN} & High; time-ordered sequential data & Implicit via temporal encoding & Low; hidden states are abstract & Dynamic trait prediction; sequential pattern learning \\
    \addlinespace
    \textbf{Encoder RNN} & High; time-series data & Stochastic encoding model uncertainty & Low; latent space hard to interpret & Latent growth pattern discovery; generative modeling \\
    \addlinespace
    \textbf{PCA}\textsuperscript{\textdagger} & Moderate; continuous multivariate data & Not explicitly handled & Low; PCs are linear combinations without direct meaning & Dimensionality reduction; exploratory trait analysis \\
    \addlinespace
    \textbf{State-Space Model (SSM)} & Moderate; time-series with latent dynamics & Explicit modeling of process and observation noise & Moderate; states interpretable if well-defined & Latent process modeling; sensor fusion; phenotype control \\
    \bottomrule
    \end{tabularx}
    \begin{tablenotes}
        \item[\textdagger] \small{Not probabilistic by default, but uncertainty can be modeled using Bayesian or variational extensions.}
        \end{tablenotes}
    \end{threeparttable}
\end{table}

\pagebreak

%%%%%% 2D/3D Structured Representations through Time %%%%%%

\section{2D/3D Structured Representations through Time}
\label{sec:structured}
\begin{figure}[h]
\centering
\includegraphics[width=.7\textwidth]{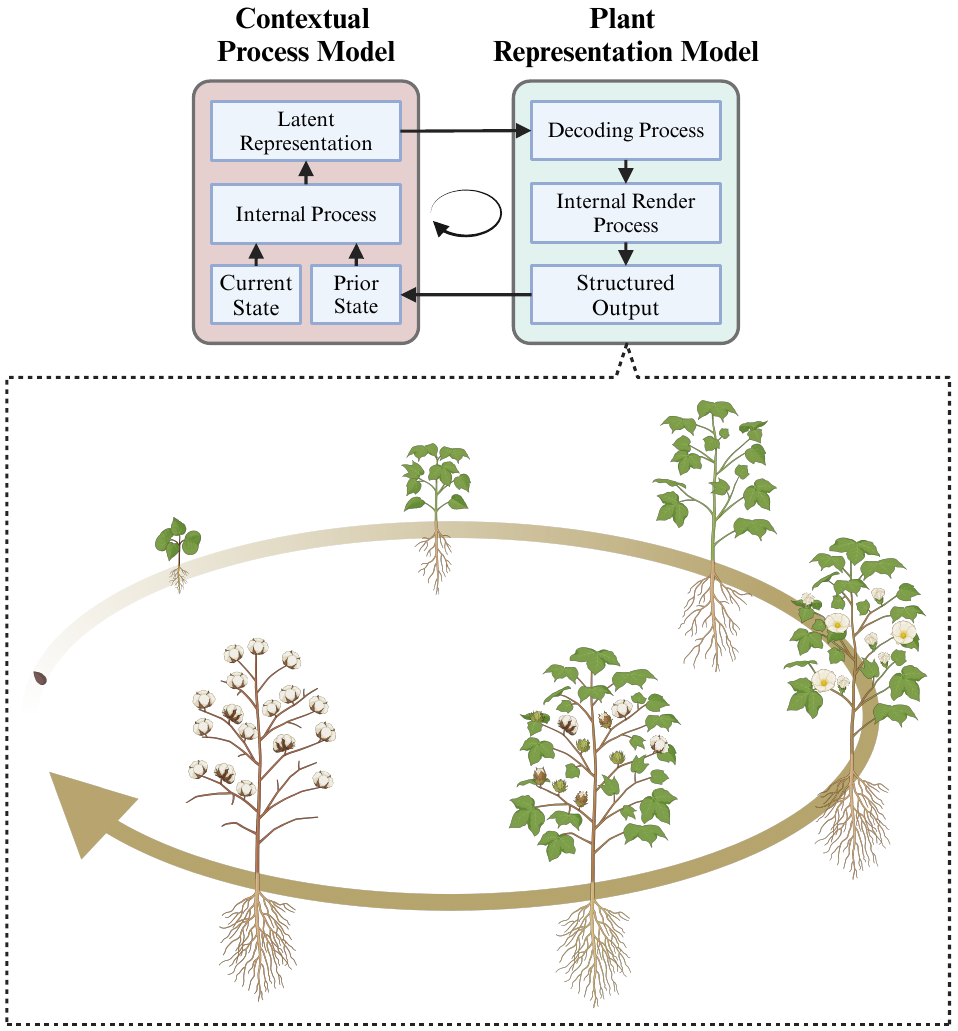}
\caption{High-level conceptual overview of a context driven simulation framework for structured output representation through time.\label{fig:Structured_render}}
\end{figure}

From low-dimensional scalar traits, we move toward representing the structural elements of plant growth through higher order data structures, such as 2D images and 3D point clouds. This is particularly useful for describing the development of plant structures through time, providing a graphical representation and visual insights to plant dynamics that are normally abstracted away with vectorized quantitative traits. The spatiotemporal simulation of these attributes offers a holistic assessment of strategies exhibited by the various plant behaviors based on contextual stimuli or signals. In this section, we take the foundation of pattern recognition techniques explored in Section \ref{sec:modeling_paradigm} and apply them toward tasks that represent plant morphologies as structured outputs. This typically involves the addition of some rendering process which defines our plant representation model and is driven by a contextual process (Figure \ref{fig:Structured_render}). As such, we explore the various works on the 2D or 3D structured plant representations, as well as works that attempt to model the entire context driven spatiotemporal pipeline for simulating plant growth.

\subsection{Functional-Structural Plant Models}
Functional-structural plant models (FSPM) represent a class of plant representation models that construct the structural and geometric elements of a plant from its mechanistic processes \cite{Godin2005,10.1093/jxb/erp345}. Particularly, these models capture the interaction between the plant's physiological processes, such as photosynthesis, water transport, and nutrient allocation, and its development of plant organs such as leaves, branches, and roots over time \cite{10.1093/aob/mcx144,Evers2016,Xiujuan2024,doi:10.1073/pnas.1410238111}. By coupling these two aspects, FSPMs provide a comprehensive view of plant behavior in response to various dependent parameters, allowing for the visualization of resource allocation and its effect on growth patterns. More recently, empirical methods have been integrated to the development process of FSPMs to establish the relationships between features extracted from HTPP data \cite{doi:10.34133/plantphenomics.0127}. As such, an emergence of hybrid data-driven, mechanistic FSPMs have been applied towards plant growth simulations.

\subsubsection{Process-driven Differential Growth Models}
One application of these FSPMs is to capture the morphological developments within plant organs through differential growth models. The development and emergence of complex 3D morphologies within plant structures through time can be represented through numerical models, such as finite element methods (FEM), to describe differential growth across localized regions. The basis of the mechanistic models that determine these differentials are typically derived through experimental observations, utilizing image-processing techniques such as optical flow to obtain growth parameters from HTPP data \cite{10.3389/fpls.2019.00227}. Changes are mapped spatially and reconstructed from the distribution of conditions which influence the final shape of plant organs. One example of such modeling is demonstrated in Huang et al. \cite{huang2018differential}, where biomechanical conditions are observed to determine the shape formations of leaves and petals. In this work, the modeling of growth differentials can be used to describe and reconstruct twisting, helical twisting, and edge-waving configurations within leaf structures through FEM simulations.

More often, this approach for plant structural modeling is applied toward developing fate and growth maps of cells within individual plant organs to simulate localized development \cite{Mosca2018}. For instance, in the work done by Kierzkowski et al. \cite{KIERZKOWSKI20191405}, the dynamics of genomic-proteomics pathways of two homeobox genes and their effects on promoting, inhibiting, and differentiating plant cells are modeled through spatial growth differentials using FEM. This phenomena is modeled through time and driven by the localized stimuli of plant hormone, namely auxin, concentrations. In this case, this model is applied toward developing a deep understanding of the causal effects of these specific signal pathways through empirical methods by mapping signals through time.

\subsubsection{Rule-based Lindenmayer systems}

\begin{figure}[H]
\centering
\includegraphics[width=\textwidth]{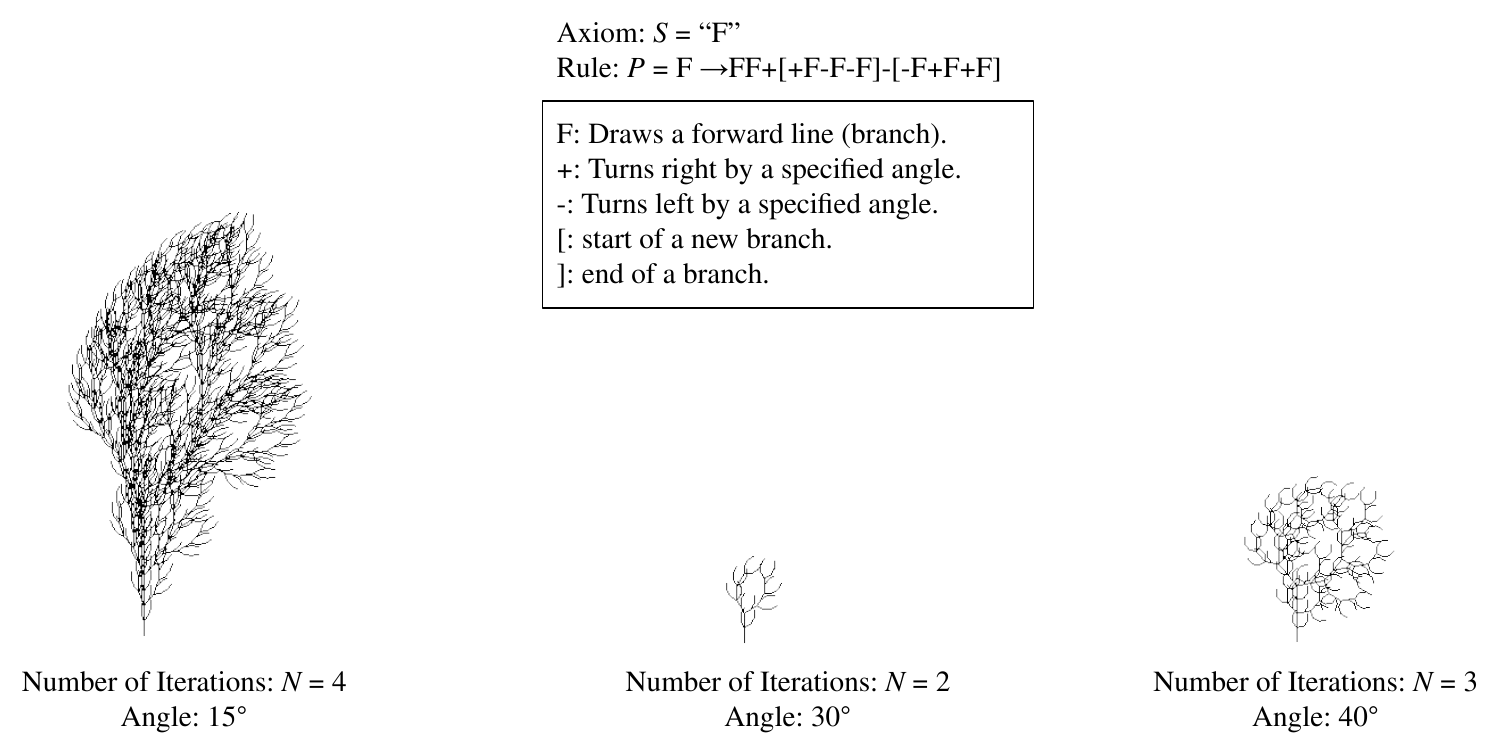}
\caption{Demonstration of plant-like structures generated using L-Systems with varying parameters. The axiom $S = \texttt{"F"}$ and production rule $P$ are iterated $N$ times with specified angles. Examples show branching patterns for $N = 2, 3, 4$ and angles $15^\circ, 30^\circ, 40^\circ$, demonstrating the influence of parameterized rules on the rendering process. The symbols represent drawing actions.\label{fig:branch_lsystem_render}}
\end{figure}

Another common implementation of FSPMs involves the application of a rule-based modeling paradigm known as the Lindenmayer systems (L-systems). L-systems define a formal set of recursive rules that specify the configuration of plant structures based on a string of symbolic characters. These rules guide the growth and branching patterns of simulated plants through iterative processes that aim to generalize and represent the realistic biological development and propagation of plant structures \cite{LINDENMAYER1968280,prusinkiewicz2012algorithmic,Prusinkiewicz2018}. By repeatedly applying these recursive rules, L-systems can simulate natural variations and intricate details found in plant morphology, making them a powerful tool in computational botany and generative modeling. 
A basic example is given by Algorithm \ref{algo:lsystem}, which establishes a generalized execution of a simple L-system, where Figure \ref{fig:branch_lsystem_render} shows the visual rendering of a rule set commonly used for describing a simple branching pattern in 2D.

\begin{algorithm}[H]
\scriptsize
\caption{Generate L-System String}\label{algo:lsystem}
\begin{algorithmic}[1]
\Require Axiom (initial string) $S$, number of iterations $N$, set of production rules $P$
\Ensure Final string after applying rules
\State Initialize $current \Leftarrow S$
\For{$i = 1$ to $N$} 
    \State $next \Leftarrow$ empty string
    \For{each character $c$ in $current$}
        \If{there is a rule $c \rightarrow replacement$ in $P$}
            \State Append $replacement$ to $next$
        \Else
            \State Append $c$ to $next$
        \EndIf
    \EndFor
    \State $current \Leftarrow next$
\EndFor
\State \Return $current$ 
\end{algorithmic}
\end{algorithm}

\pagebreak

This algorithm can be expanded to include parameterized production rules that encapsulate complex relationships at each iteration of the simulated plant branching. For example, various works have developed production rules that dictate tree organ formation based on a source-sink dynamic between existing plant organs at previous iterations, which is then subsequently rendered for the 3D simulation of various tree cultivars \cite{LPeaches2005,dejong2015almond}. More recent work has expanded on production rules using sink-source dynamics by incorporating context-driven systems such as environmental and management conditions, soil nutrient availability, and distribution to improve the dynamic tree simulation process \cite{LPeaches2022,9231935,YANG2024122}. 

Furthermore, works such as Hitti et al. \cite{HITTI2024108613} and Mu{\ss}mann et al. \cite{10.1162/isal_a_00718} introduce an adaptation of the L-systems paradigm that captures the dynamic development of plant structures using agent-based interactions. Plant structures, represented as agents, can independently interact with various stimuli. Within such systems, agent-based feedback mechanisms, such as reinforcement learning, can be used to train model growth patterns based on external stimuli.

\subsection{Computer Vision Approach to Plant Modeling}
\label{sec:CV}
FSPMs typically rely on procedural methods for the visual rendering process of the plant simulation pipeline (Figure \ref{fig:Structured_render}). Although the outcomes of such models are interpretable and result in visually consistent structures, the outputs of such methods are often derived from deterministic solutions that lack stochastic variability. Such solutions tend to drift from biological realism, particularly when simulating complex interactions or environmental responses, requiring the integration of experiment data to drive realistic outputs \cite{10.1093/jxb/erp345}. Alternatively, some initial efforts have been made towards a novel approach using data-driven CV for rendering the structural elements of the simulation task. 

For instance, the task of plant growth modeling can be structured to fall within one of two classic problems found within the domain of predictive CV and video analysis; frame-to-frame translation and frame synthesis. Within this perspective, the structural elements of plant growth can be represented as images of a temporal scene, or frames, where the goal of our methods is to predict the pixel-wise configuration of these frames, such that the corresponding scene describes the realistic spatiotemporal development of plant growth. These methods do not rely on procedural rendering routines using pre-rendered graphics. Rather, the visual rendering process becomes a CV-based image translation task driven by empirical methods for predicting frames through features learned from high dimensional 2D or 3D HTPP data. We explore some early works which utilize such methods for predictive pattern recognition.

\begin{figure}[t]
\centering
\includegraphics[width=\textwidth]{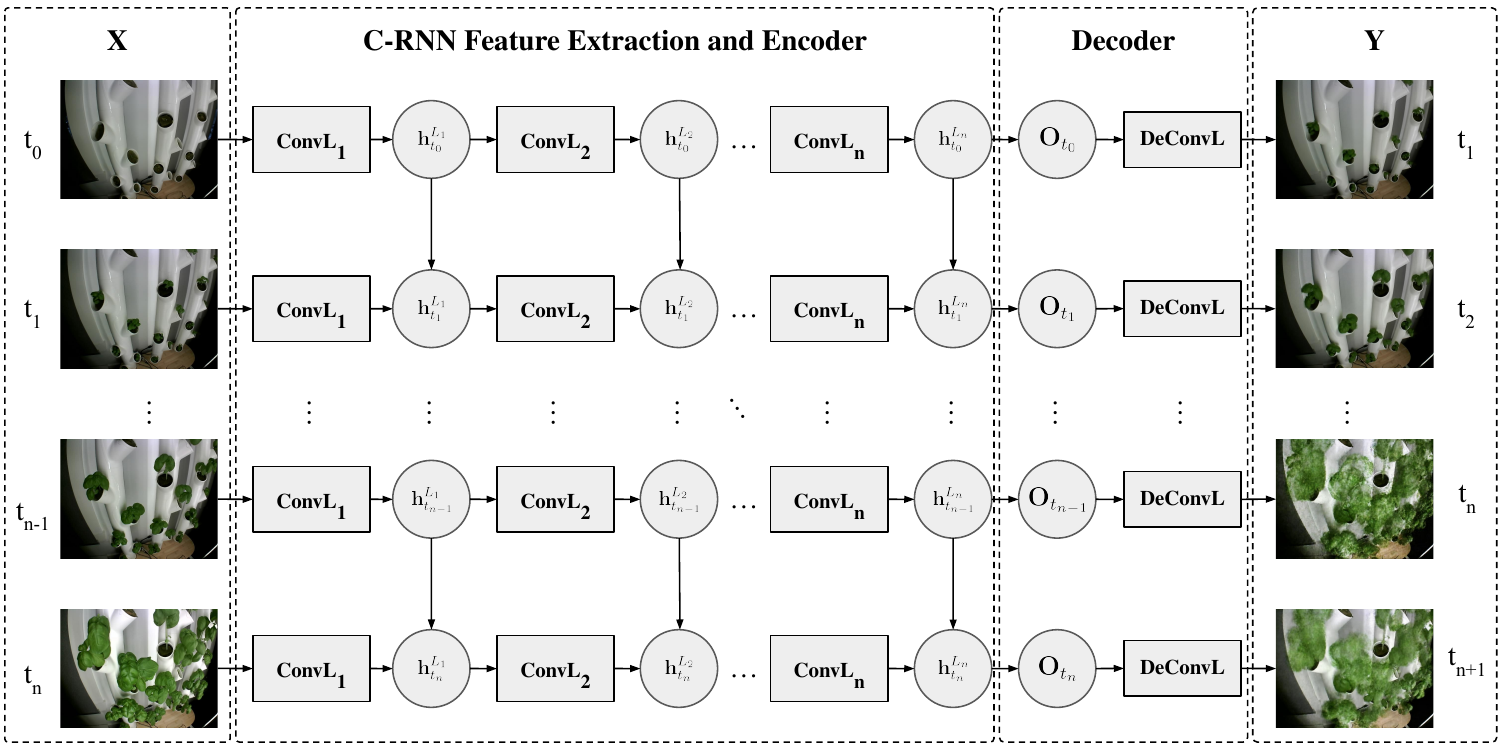}
\caption{Directed graph model (unfolded) of a Convolutional Recurrent Neural Network (CRNN) implementation for predicting the next sequence in a plant scene. The model includes convolutional layers (ConvL1, ConvL2, ..., ConvLn) for feature extraction and encoding, followed by a recurrent layer for temporal processing of sequential frame inputs,$X$, at $t_0, t_1, ..., t_n$. A decoder module with deconvolution layers (DeConvL) reconstructs the output frames, $Y$, at $t_{n+1}$ based on learned spatiotemporal features.\label{fig:crnn}}
\end{figure}

\subsubsection{Next Frame Prediction} 
\label{sec:nextframeprediction}
Some early works on CV-based plant growth forecasting include Sakurai et al. \cite{sakurai2019plant} and Wang et al. \cite{agronomy12092213}, which have proposed the adoption of convolutional recurrent neural networks (CRNN) for the task of image-to-image translation to predict the next frame of a plant growth scene given a set of prior frames from the same scene. This approach expands on the sequence-based prediction techniques described in Section \ref{rnn} to assess spatial features from consecutive images. This is done by incorporating convolutional layers within a RNN framework to extract and encode features from a set of input images, $X$, to a transitional hidden state $h$. The rendering of this hidden state is described by a decoding step which consists of an up-sampling process of $h$ through deconvolutional layers. The typical generic implementation of the CRNN is shown in Figure \ref{fig:crnn}. This approach combines pattern recognition techniques for forecasting plant parameters using RNNs with the visual rendering of plant structures through image translation methods. While CRNNs allow for spatial features to be captured through time, the deterministic mapping through convolutional and deconvolutional layers often result in artifacts and ambiguities within structures of the output images \cite{kingma2013auto}.

To mitigate problems in image quality, recent work has focused on improving the image generation component of model architectures, which correspond to the decoder of a CRNN. Yasrab et al. \cite{rs13030331} proposes the use of a Generative Adversarial Network (GAN)-based network architectures for improved realistic image synthesis. Other works such as Kim et al. \cite{9750078} propose multi-step networks that separate various components of the image generation task, such as shape estimation and image reconstruction, utilizing a transformer-based network for spatial pattern recognition. Although these approaches improve image generation quality, they do not address the challenges of ambiguous structures caused by a deterministic mappings. CV-based plant forecasting is an ill-posed problem, in which, there exists non-unique outcomes of a plants growth trajectory given the same initial set of frames. This is particularly evident under diverging environmental conditions. Moreover, predictions from the current models are made solely on the previous frames of the scene and further contribute to the prevalence of ambiguities during image translation.

\begin{figure}[t]
    \centering
    \includegraphics[width=\textwidth]{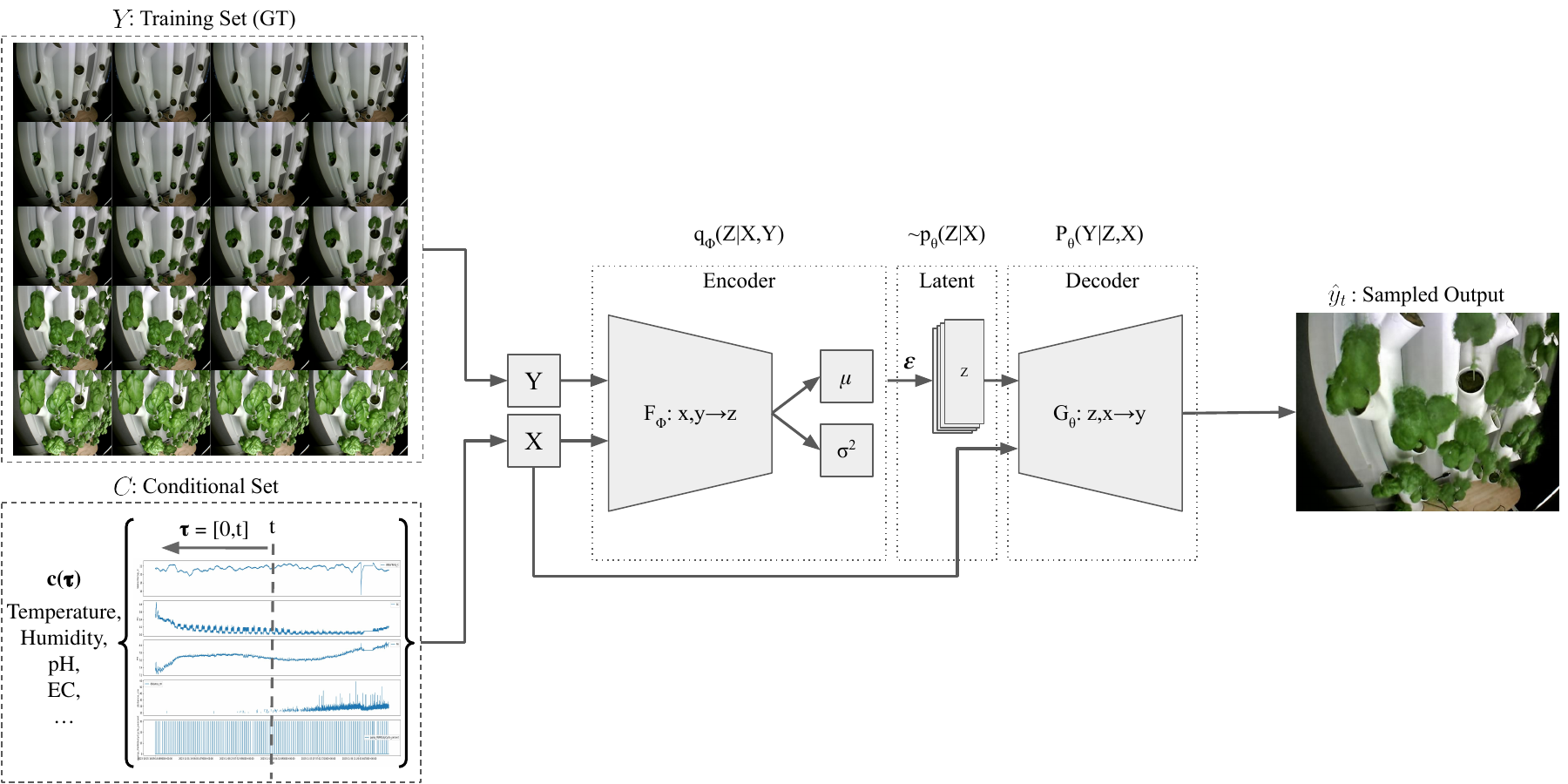}
    \caption{Conditional generative model architecture for guided plant scene simulation. The framework includes a latent encoder $F_\Phi(x, y \rightarrow z)$ that maps input data ($X, Y$) to a latent variable $Z$, and a decoder $G_\Theta(z, x \rightarrow y)$ that reconstructs ground truth (GT) outputs, $Y$. The latent space is defined by $\mu$ and $\sigma^2$ and incorporates auxiliary stochastic noise $\epsilon$. Conditional vector, $X$, is directed by environmental parameters, $c(\tau)$, such as temperature, humidity, pH, and EC, over a time period $\tau = [0, t]$. The architecture allows for generating sampled outputs, $\hat{y}$, guided by $X$.\label{fig:cgm}}
    \end{figure}

\subsubsection{Frame Synthesis Through Conditional Generative Models}
\label{generative process}

\begin{figure}[t]
    \centering
    \includegraphics[width=\textwidth]{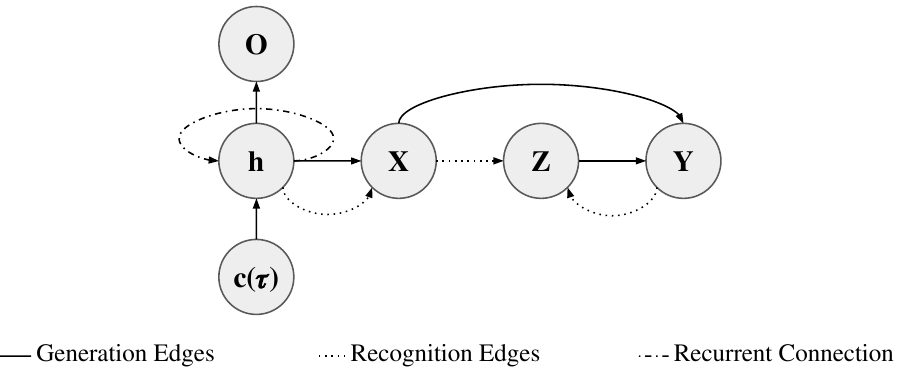}
    \caption{Directed graph model (folded) of a sequence-informed method for incorporating compounding environmental feedback to the conditional generative model (CGM). Environmental parameters, $c(\tau)$, are sequentially processed through an RNN encoder to obtain conditional vector, $X$. $X$ guides the generation and latent encoding process of the CGM to produce output frames, $Y$. The generation edges correspond to the synthesis of outputs frames, $Y$, while the recognition edges correspond the process of learning the probability distributions of latent variable, $Z$. The recurrent connections allow the system to integrate sequential information for dynamic plant growth predictions. \label{fig:sipgs}}
    \end{figure}

A probabilistic approach can be taken to regularize the plant growth modeling problem and to guide the prediction process by incorporating stochastic elements caused by dynamic growing conditions. In such methods, a generative model is employed for the task of frame synthesis to predict the expected frames of the plant growth scene. While generative models are typically used as data augmentation techniques for enhancing sparse agricultural datasets \cite{doi:10.34133/plantphenomics.0041, LU2022107208}, few works have explored the use of conditional generative models (CGM) for guided image synthesis. A probabilistic approach using a CGM is illustrated by a generic conditional variational autoencoder (CVAE) shown in figure \ref{fig:cgm}. Instead of performing direct deterministic mappings between inputs and output frames, the objective of the CVAE is to learn a prior distribution, $p_\theta(z|x)$, where $z\sim p_\theta(z|x)$ is the latent representation of scene at a given point in time, $t$, directed by some conditional vector $x$. The prior distribution can be sampled such that it incorporates stochasticity through auxiliary variable, $\epsilon$. This process is known as latent sampling and it guides the generation process by  randomly sampling from an underlying latent distribution that captures the hidden structure of the data. The network structures explored within the domain of plant growth modeling typically consists of a probabilistic latent variable encoder and a deterministic decoder to render output frames, $\hat{y} \sim p_\theta(y|z,x)$.

Early works, including Miranda et al. \cite{9956115} and Drees et al. \cite{DREES2021106415}, propose the adoption of a conditional GAN (cGAN), which introduces a discriminator with adversarial routines to the generic CVAE structure, for synthesizing field images and forecasting the state of the crop canopy. The conditional vector consists of fixed soil treatment embeddings set at the start of the growing season. Forecasts are made at discrete points in time driven by the initial states of the treatment conditions. These works have demonstrated SOTA guided image generation for realistic image synthesis with variability in outputs defined by stochastic elements, or \textit{``factors of variation'' (FOV)}. The frames synthesized from such models can be used in conjunction with image-based feature extraction models to provide quantitative estimation of biomass for a complete simulation process \cite{10.1007/978-3-031-16788-1_29}.

In these works, predictions are guided by a conditional vector comprised of treatment conditions, a prior frame state, and a time factor. The treatment embeddings are fixed at a particular point in time, typically set at the start of experimentation, and do not account for dynamic changes in conditions. For spatiotemporal simulation and trajectory estimation, the guidance of predictions without feedback can lead to larger variability when diverging from initial conditions due to the accumulated effects of compounding factors, including environmental conditions \cite{Drees_2024}. As such, this approach may be applicable for simulating images of end of harvest canopies under varying treatment conditions at a discrete point in time in the future, but lacks the dynamics for real-time continuous plant growth predictions useful for real world applications such as anomaly detection and intervention practices.

Debbagh et al. \cite{Debbagh2024SIPGS} proposes a sequence-informed approach which introduces dynamic and compounding environmental conditions as feedback to a CGM, guiding a continuous simulation process. In this work, the conditional vector, $x$, which guides the image generation of the CGM, is directed by a configuration of random variables, $c$, that describe dynamic factors that contribute to the development of the plants within the scene (i.e. environmental parameters). $x$ is determined by encoding a sequence of these environmental parameters from the start of the simulation up to the point of inference, $c(\tau)$, where $\tau = [0,t]$. The encoding process consists of an RNN encoder, which allows the model to learn the compounding and sequential relationships between environmental conditions as a vector representation. The dependencies between each model component is illustrated by the directed graph model shown in Figure \ref{fig:sipgs}. This approach provides a framework that further regularizes the image generation task through dynamic context, allowing for the realistic synthesis of consecutive frames.

\subsection{Comparisons of Structured Output Representations Through Time}

Table~\ref{tab:structured_models} provides a comparative summary of the structured modeling paradigms discussed in this section for simulating plant morphologies through time. Each method represents a trade-off between data needs, realism, and interpretability.

Mechanistic FSPMs such as differential growth simulations based on FEM rely on high-resolution imaging and cell-level observations to reconstruct localized growth processes. These models are highly interpretable, driven by physical and biological insights, but are typically deterministic and do not formally account for uncertainty. 
L-systems offer interpretable and symbolic modeling of branching structures using production rules. While L-systems are not probabilistic by default, stochastic extensions and agent-based variants allow for the incorporation of environmental interaction and feedback, enabling more dynamic simulations.

% \pagebreak

CV-based approaches introduce data-driven alternatives for visualizing plant growth. CRNNs are used to capture sequential spatial features from image sequences but suffer from low interpretability and deterministic output mappings, while CGMs, including the sequence-informed CVAE and cGANs, address these challenges by introducing latent variable modeling and sampling to represent uncertainty and generate more biologically plausible outputs.

\begin{table}[h]
    \centering
    \begin{threeparttable}
    \caption{Comparison of Structured Models for Simulating Plant Morphologies Through Time}
    \label{tab:structured_models}
    \scriptsize
    \begin{tabularx}{\textwidth}{@{\extracolsep{\fill}}L{0.14\textwidth}L{0.18\textwidth}L{0.21\textwidth}L{0.17\textwidth}Y@{}}
    \toprule
    \textbf{Model} & \textbf{Data Requirements} & \textbf{Uncertainty Handling} & \textbf{Interpretability} & \textbf{Application} \\
    \midrule
    % \textbf{FSPM (Mechanistic)} & Moderate; physiological and environmental parameters & Deterministic; no formal uncertainty modeling & High; process-based and interpretable & Resource allocation, physiological responses, whole-plant simulation \\
    
    % \addlinespace
    \textbf{Differential Growth (FEM)} & High; cell-level growth data, imaging-based flow analysis & Implicit in parameter variability; deterministic model & High; spatial and physical dynamics interpretable & Leaf morphogenesis, biomechanical growth simulations \\
    
    \addlinespace
    \textbf{L-System}\textsuperscript{\textdagger} & Low to moderate; rule sets and symbolic encodings & Not explicitly handled & High; rules directly map to morphology & Branching patterns, 2D/3D growth simulations, generative modeling \\
    
    \addlinespace
    \textbf{Agent-based L-System} & High; interaction data with agents/environment & Stochasticity and feedback via agent logic; tunable & Moderate to high; emergent from explicit interactions & Context-aware growth modeling; reinforcement learning driven growth \\
    
    \addlinespace
    \textbf{CRNN (Next Frame Prediction)} & High; sequential image data of plant growth & Implicit via temporal encoding & Low to moderate; convolution and hidden state opacity & Visual trait prediction; spatiotemporal structure forecasting \\
    
    \addlinespace
    \textbf{Conditional Generative Models (CVAE, cGAN)} & High; image pairs with metadata or environmental labels & Explicit; latent distribution sampling (KL divergence, adversarial) & Moderate; constrained latent space enables partial interpretation & Visual scene generation; growth under static treatment conditions \\
    
    % \addlinespace
    % \textbf{Sequence-Informed Conditional Generative Models} & Very high; sequential environmental conditions and image frames & Explicit; feedback-aware latent modeling and sampling & Moderate; latent dynamics linked to environmental context & Continuous forecasting; closed-loop simulation under dynamic conditions \\
    \bottomrule
    \end{tabularx}
    \begin{tablenotes}
        \item[\textdagger] \small{Not probabilistic by default, but uncertainty can be modeled using Bayesian or variational extensions.}
        \end{tablenotes}
    \end{threeparttable}
\end{table}

%%%%%%%%%%%%%%%%%%%%%%% Perspectives

\section{Perspectives}
\label{sec:perspectives}
\subsection{Data-Driven vs Experiment-Driven Approaches}
Plant growth modeling has traditionally relied on experiment-driven approaches, where predefined hypotheses are tested using carefully controlled experiments. 
These methods provide valuable insights into specific biological processes but are often time-consuming, resource-intensive, and limited in scalability \cite{akhavizadegan2021time,suliansyah2023literature,kephe2021challenges}. 
By contrast, data-driven and model-centric approaches leverage large datasets and machine learning techniques to capture patterns and relationships that might not be immediately apparent within an experiment-driven framework \cite{roscher2023datacentricdigitalagricultureperspective}. The works presented in this review demonstrate a shift towards the latter, as challenges with dynamic updating and long-term predictions are prevalent in simple regression methods as described in Section \ref{pitfalls of regression}. 

The shift towards 2D image and 3D point cloud data for describing plant phenemena, as with HTPP datasets, requires model frameworks designed for interpreting and processing higher order data-structures \cite{harandi2023make,tpj.17053}. 
This inherently moves prospective research towards the CV-based models described in Sections \ref{sec:nextframeprediction} and \ref{generative process}. However, the advantages of these data-driven approaches come with the limitation of reduced interpretability and a lack of the biological grounding inherent in experiment-driven methods \cite{mostafa2023explainable}. 
Model frameworks such as FSPMs present potential opportunities for incorporating knowledge-based approaches for structured representation that is highly controllable, but is completely reliant on experiment-driven approaches for developing knowledge-based rule sets.

Future research in this domain should prioritize the development of hybrid approaches that integrate the strengths of both paradigms. 
Two key directions are proposed to achieve this goal. The first involves designing model architectures that incorporate knowledge-informed mechanisms, enabling the integration of biological principles and domain expertise into data-driven frameworks. 
This is demonstrated by works that aim to integrate deep learning frameworks into large scale crop models such as CERES and DSSAT \cite{shi2025deep}. The second focuses on advancing methods for decoding and interpreting opaque models, ensuring that complex representation models provide outputs that align with biological processes and facilitate deeper understanding \cite{mostafa2023explainable}.

\subsection{Dataset Gap}

The shift toward data-driven approaches has placed datasets at the center of plant growth modeling and predictive frameworks. 
Large-scale, high-quality datasets are necessary for training DL models and validating their predictions \cite{xu2023embracing,kiab301}. 
However, the effectiveness of these models are dependent on the availability, diversity, and representativeness of the data \cite{LI2023108412}. 
Despite advancements in HTPP platforms, significant gaps remain in the scope, quality, and accessibility of existing datasets. 

First, many datasets focus on single time points or static snapshots, which fail to capture the dynamic growth trajectories of plants over time. In such cases, works such as Zhu et al. \cite{zhu2020analysing} aim to simulate and analyze the physiological development of soybean plants using interpolated 3D reconstructions of discrete snapshots. 
Existing publicly available datasets that capture longitudinal data, compiled together and shown in Table \ref{dataset list}, are  scarce and limited in sample frequency. 
This is particularly evident for datasets that capture multiple modalities of data including 2D images and 3D point clouds along with environmental parameters. 
Second, the diversity of crop species represented by current datasets in Table \ref{dataset list} and HTTP platforms is limited, with a heavy emphasis on staple crops such as maize (\textit{Zea mays}) and soybean (\textit{Glycine max}) \cite{s141120078}. 
This narrow scope restricts the ability of models to generalize across plant species with differing morphologies, such as root vegetables or perennials.
Finally, the lack of consolidation of datasets across different research groups and platforms creates significant barriers to broader adoption and cross-comparative studies. Inconsistent formats, annotation practices, and metadata standards hinder the ability to combine datasets, thereby limiting the diversity and scale necessary for developing generalizable models \cite{baab028}.

To address these gaps, several considerations for developing future datasets are proposed. First, works should aim to apply standardization and open-access practices during dataset development to enable reproducibility and comparability across studies through benchmarking. High-frequency sampling should also be prioritized to produce longitudinal datasets that span entire growth cycles, improving the ability to model and predict temporal dynamics accurately. Furthermore, the development of multi-modal datasets, combining 2D images, 3D point cloud data, and environmental sensor measurements is suggested for capturing a holistic view of plant-environment interactions. Finally, integrating domain knowledge into dataset design to ensure that datasets not only capture observable traits but also reflect underlying biological processes. Such biologically informed datasets will enable models to generate predictions that are both empirically and biologically meaningful \cite{schramowski2020making}.

\subsection{Practical Applications}
The recent advancements in pattern recognition techniques explored in this review have significant implications for practical applications. While traditional applications include crop yield forecasting and climate simulations, these novel approaches expand the scope to include controlled environment systems, field crop systems, and plant genomics and phenomics research. These advancements allow for precise detection, simulation, and representation of plant growth processes in complex environments.

\subsubsection{Anomaly Detection}
Anomaly detection and early interventions practices during crop development benefit from predictive pattern recognitions techniques that incorporate dynamic feedback from environmental growing conditions. 
The focus on the development of models that capture long-term temporal dependencies through sequence-based methods, such as transformers and RNNs, provides an effective approach to identifying deviations in plant growth trajectories. 
For instance, temporal anomalies such as stunted growth or disease occurrence can be detected in advance by analyzing sequential data from real-time HTPP platform measurements \cite{10.3389/fpls.2022.890563}. 
Moreover, the incorporation of multi-modal approaches to combine imaging data with contextual environmental parameters presents a more comprehensive assessment of plant health. 
Models such as multi-variate transformers have been employed to improve the precision of anomaly detection in time-series data \cite{tuli2022tranad,xu2021anomaly}. Such techniques are particularly useful for early intervention, mitigating potential yield losses and optimizing resource use.

\subsubsection{Plant-Environment Simulation}
Simulation frameworks informed by generative and probabilistic models improve the ability to predict plant growth under varying environmental conditions. 
Models like the CGMs presented in Section \ref{generative process} enable the synthesis of realistic growth trajectories influenced by the spatial and dynamic growing condition, including the distribution of light, nutrient conditions, temperature, and CO\textsubscript{2} \cite{Debbagh2024SIPGS,DREES2021106415}. 
This provides \textit{in silico} methods to inform the structural development process of a controlled plant system and to develop strategies prior to conducting in-field operations. These frameworks allow for resource optimization in controlled environments and improve the forecasting of long-term plant responses to changing environments.

\subsubsection{Digital Twinning and Representation}
Digital twinning is the virtual replication of a physical system, maintained through continuous data exchange, simulation, and feedback. 
Within plant phenomics, digital twinning is limited and typically consists of methods for 3D reconstruction of plant structures taken at sparse static snapshots within its growth cycle. 
This is often due to limitations in imaging techniques for capturing frequent 3D spatial data during its growth and challenges in continuous collection of auxiliary data \cite{harandi2023make,LI2022106712}. 
The structured representation methods explored within this review have implications on expanding this static framework to capture a continuous virtual replica of a plant system during its entire growth cycle. 
The continuous frame synthesis techniques explored in Debbagh et al. \cite{Debbagh2024SIPGS}, allow for the interpolation of structured outputs between points of time. 
Moreover, FSPMs such as L-systems also demonstrate a potential for continuous state estimation by basing structural development on biological rules as a method that does not rely on continuous data collection \cite{MITSANIS2024108733}. 
These methods present an opportunity to generate digital plant structures with the guidance of few snapshots of plants. 
Moreover, the divergence of these plant structures from the snapshots can be explored through simulated growing conditions. 

\section{Conclusion}

In this review, we provide a comprehensive examination of recent works and explore prospects towards advancing predictive pattern recognition techniques within plant phenomics research. These works cover various approaches that address challenges in processing and interpreting various representations of spatiotemporal plant data, which include longitudinal scalar growth traits and structured 2D images and 3D point cloud data. Within these works, novel modeling paradigms, which take advantage of recent advancements in computational systems, HTPP data collection and data fusion techniques, have presented significant improvements to the precision and accuracy of models for long-term forecasting and other simulation tasks. Notably, NN-based representation models and CGMs present a shift toward a data-driven approach to predictive pattern recognition that not only allows for dynamic updating for guiding predictions, but also allows for the synthesis of structured representation for spatial understanding of outputs.

\pagebreak

Additionally, this review offers a perspective for opportunities to advance works within in this field. This includes perspectives on the integration of domain knowledge with data-driven approaches, and the emphasis on developing comprehensive, standardized datasets. Furthermore, we suggest practical applications within various domains to be explored in future works. The collection of these works and our perspective aims to consolidate and expand the understanding of spatiotemporal modeling frameworks and their role in advancing plant phenomics research and applications.

%%%%%%%%%%%%%%%%%%%%%%%%%%%%%%%%%%%%%%%%%%%%%%%%%%%%%%%%%%%%%%%%%%%%%%%%%%%%%%%%%%%%%%%%%%%%%%%%%%%%%%%%%%%%%%%%%%%%%%%%%%%%%%%%%

\section*{Acknowledgments}
\subsection*{General} 
Special thanks to Anita Parmar, Ollivier Dyens, and the Building 21 community for their valuable contributions through engaging and thought-provoking discussions.
\subsection*{Author Contributions} 
M. Debbagh drafted the manuscript, prepared the figures and tables, and incorporated feedback from co-authors. S. Sun and M. Lefsrud reviewed and edited the manuscript, offering feedback to improve clarity and impact.

\subsection*{Funding}
This work was supported by the Natural Sciences and Engineering Research Council of Canada (RGPIN-2021-03266, RGPIN-2021-03410); and Gardyn through the Mitacs Accelerate program (IT16220).

\subsection*{Tools} 
Figure \ref{modeling paradigm} (https://BioRender.com/y80s089) and Figure \ref{fig:Structured_render} (https://BioRender.com/y69z168) were created in BioRender.com.

\subsection*{Conflicts of Interest}
The authors declare that there is no conflict of interest regarding the publication of this article.

% \printbibliography
\pagebreak

\bibliographystyle{unsrtnat} 
\bibliography{ref} 

\end{document}